\documentclass[10pt,twocolumn,letterpaper]{article}

\usepackage{cvpr}
\usepackage{rotating}
\usepackage{times}
\usepackage{epsfig}
\usepackage{graphicx}
\usepackage{amsmath}
\usepackage{amssymb}
\usepackage{amsfonts}
\usepackage{empheq}
\usepackage{float}
\usepackage{framed, color}
\usepackage{xcolor}
\usepackage{graphics}
\usepackage{graphicx}
\usepackage[]{graphicx} 
\usepackage{epsfig} 
\usepackage{subfigure}
\usepackage{algorithm}
\usepackage{algorithmic}
\usepackage{stmaryrd}
\usepackage[mathscr]{eucal}
\usepackage{lineno}
\usepackage{color}
\usepackage{filecontents}
\usepackage{subfigure}
\usepackage{multirow}
\usepackage{filecontents}
\usepackage{verbatim} 
\usepackage{tabularx}
\usepackage{lineno}
\usepackage{setspace}
\usepackage{multirow}
\usepackage{textcomp,booktabs}
\usepackage{caption}

\def\I{{\bf I}}

\def\w{{\bf w}}
\def\0{{\bf 0}}
\def\1{{\bf 1}}

\newcommand*{\colorboxed}{}
\def\colorboxed#1#{%
	  \colorboxedAux{#1}%
}

\newcommand*{\colorboxedAux}[3]{%
	\begingroup
	\colorlet{cb@saved}{.}%
	\color#1{#2}%
	\boxed{%
		\color{cb@saved}%
		#3%
	}%
	\endgroup
}

\usepackage[pagebackref=true,breaklinks=true,letterpaper=true,colorlinks,bookmarks=false]{hyperref}

\cvprfinalcopy 

\def\cvprPaperID{1601} 

\ifcvprfinal\fi
\begin{document}

\title{Image Reconstruction with Predictive Filter Flow}

\author{Shu Kong, Charless Fowlkes\\
Dept. of Computer Science, University of California, Irvine\\
\texttt{\{skong2, fowlkes\}@ics.uci.edu } \\ \\
  \ [\href{https://www.ics.uci.edu/~skong2/pff.html}{Project Page}],
[\href{https://github.com/aimerykong/predictive-filter-flow}{Github}],
[\href{http://www.ics.uci.edu/~skong2/slides/predictive_filter_flow_slides.pdf}{Slides}],
[\href{http://www.ics.uci.edu/~skong2/slides/predictive_filter_flow_poster.pdf}{Poster}]
}

\maketitle

\begin{abstract}
We propose a simple, interpretable framework for solving a wide range of image
reconstruction problems such as denoising and deconvolution.  Given a
corrupted input image, the model synthesizes a spatially varying linear filter
which, when applied to the input image, reconstructs the desired output. The
model parameters are learned using supervised or self-supervised training.
We test this model on three tasks: non-uniform motion blur removal,
lossy-compression artifact reduction and single image super resolution.  We
demonstrate that our model substantially outperforms state-of-the-art methods
on all these tasks and is significantly faster than optimization-based
approaches to deconvolution.  Unlike models that directly predict output pixel
values, the predicted filter flow is controllable and interpretable, which we
demonstrate by visualizing the space of predicted filters for different tasks.\footnote{
Due to that arxiv limits the size of files, we put
high-resolution figures, as well as a manuscript with them, in the project page.}
\end{abstract}

\section{Introduction}

\begin{figure*}[t]
    \centering
    \begin{minipage}{0.99\textwidth}
        \centering
        \includegraphics[width=1\linewidth]{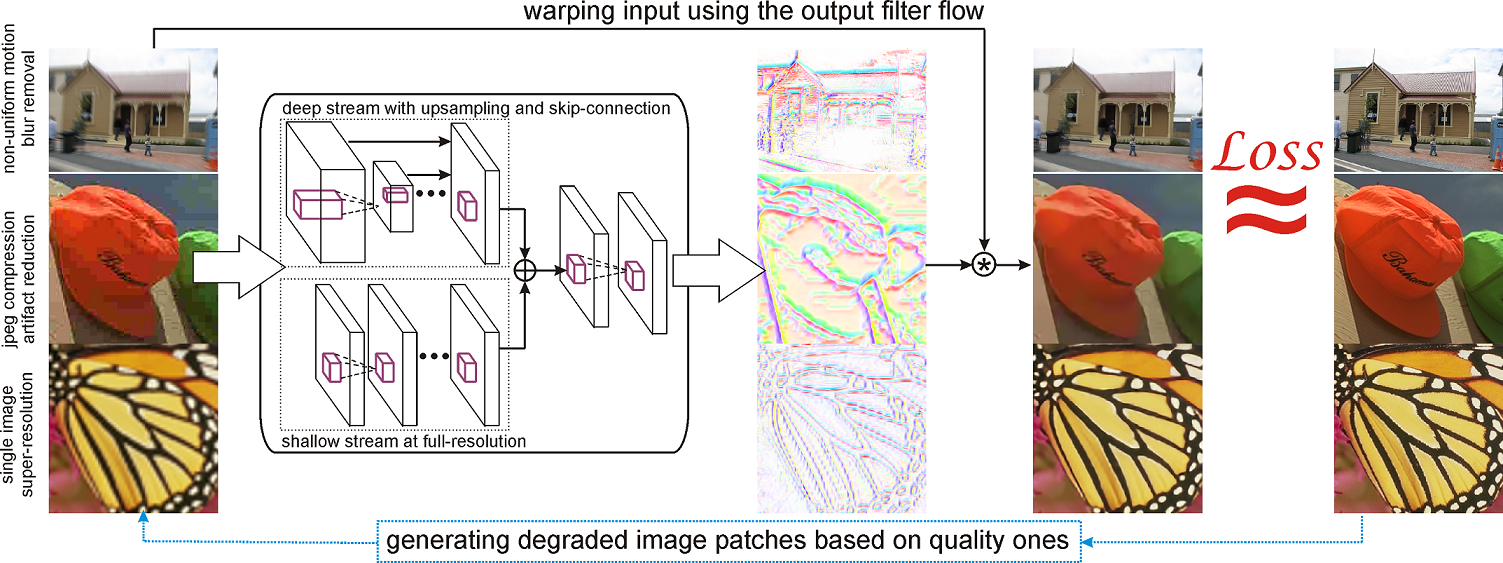}
        \captionsetup{width=0.999\textwidth}
    \end{minipage}
    \vspace{-3mm}
    \caption{\small Overview of our proposed framework for \emph{Predictive
    Filter Flow} which is readily applicable to various low-level vision
    problems, yielding state-of-the-art performance for non-uniform motion blur
    removal, compression artifact reduction and single image super-resolution.
    Given a corrupted input image, a two-stream CNN analyzes the image and
    synthesizes the weights of a spatially-varying linear filter. This filter
    is then applied to the input to produce a deblurred/denoised prediction.
    The whole framework is end-to-end trainable in a self-supervised way for
    tasks such as super-resolution where corrupted images can be generated
    automatically.  The predicted filters are easily constrained for different
    tasks and interpretable (here visualized in the center column by the mean
    flow displacement, see Fig.~\ref{fig:tSNE_filter_visualization}).}
    \label{fig:diagram}
    \vspace{-3mm}
\end{figure*}

Real-world images are seldom perfect. Practical engineering trade-offs entail
that consumer photos are often blurry due to low-light, camera shake or object
motion, limited in resolution and further degraded by image compression
artifacts introduced for the sake of affordable transmission and storage.
Scientific applications such as microscopy or astronomy, which push the
fundamental physical limitations of light, lenses and sensors, face similar
challenges. Recovering high-quality images from degraded measurements has
been a long-standing problem for image analysis and spans a range of tasks
such as blind-image deblurring ~\cite{bell1995information,kundur1996blind,
fergus2006removing,shan2008high}, compression artifact
reduction~\cite{shen1998review,list2003adaptive}, and single image
super-resolution~\cite{park2003super,yang2010image}.

%
%

Such image reconstruction tasks can be viewed mathematically as inverse
problems~\cite{tarantola2005inverse,kaipio2006statistical}, which are typically
ill-posed and massively under-constrained. Many contemporary techniques to
inverse problems have focused on regularization techniques which are amenable
to computational optimization. While such approaches are interpretable as
Bayesian estimators with particular choice of priors, they are often
computationally expensive in
practice~\cite{fergus2006removing,shan2008high,bahat2017non}.  Alternately,
data-driven methods based on training deep convolutional neural networks yield
fast inference but lack interpretability and guarantees of
robustness~\cite{sun2015learning,zhang2018residual}.  In this paper, we propose
a new framework called \emph{Predictive Filter Flow} that retains
interpretability and control over the resulting reconstruction while allowing
fast inference.  The proposed framework is directly applicable to a variety of
low-level computer vision problems involving local pixel transformations.

As the name suggests, our approach is built on the notion of filter flow
introduced by Seitz and Baker~\cite{seitz2009filter}.  In filter flow pixels in
a local neighborhood of the input image are linearly combined to reconstruct
the pixel centered at the same location in the output image.  However, unlike
convolution, the filter weights are allowed to vary from one spatial location to
the next. Filter flows are a flexible class of image transformations that can
model a wide range of imaging effects (including optical flow, lighting
changes, non-uniform blur, non-parametric distortion). The original work on
filter flow~\cite{seitz2009filter} focused on the problem of estimating an
appropriately regularized/constrained flow between a given pair of images.
This yielded convex but impractically large optimization problems (e.g., hours
of computation to compute a single flow).  Instead of solving for an optimal
filter flow, we propose to directly predict a filter flow given an input image
using a convolutional neural net (CNN) to regress the filter weights.  Using a
CNN to directly predict a well regularized solution is orders of magnitude
faster than expensive iterative optimization.

Fig.~\ref{fig:diagram} provides an illustration of our overall framework.
Instead of estimating the flow between a pair of input images, we focus on
applications where the model predicts both the flow and the transformed image.
This can be viewed as ``blind'' filter flow estimation, in analogy with blind
deconvolution.  During training, we use a loss defined over the transformed
image (rather than the predicted flow). This is closely related to so-called
self-supervised techniques that learn to predict optical flow and depth from
unlabeled video data \cite{garg2016unsupervised,godard2017unsupervised,jason2016back}.
Specifically, for the reconstruction tasks we
consider such as image super-resolution, the forward degradation process can be
easily simulated to generate a large quantity of training data without manual
collection or annotation.

The lack of interpretability in deep image-to-image regression models makes it
hard to provide guarantees of robustness in the presence of adversarial
input~\cite{kurakin2016adversarial}, and confer reliability needed for
researchers in biology and medical science~\cite{litjens2017survey}.
Predictive filter flow differs from other CNN-based approaches in this regard
since the intermediate filter flows are interpretable and
transparent~\cite{vellido2012making, doshi2017towards,Lipton18}, providing an
explicit description of how the input is transformed into output.  It is also
straightforward to inject constraints on the reconstruction (e.g., local
brightness conservation) which would be nearly impossible to guarantee for deep
image-to-image regression models.

To evaluate our model, we carry out extensive experiments on three different
low-level vision tasks, non-uniform motion blur removal, JPEG compression
artifact reduction and single image super-resolution.  We show that our model
surpasses all the state-of-the-art methods on all the three tasks. We also
visualize the predicted filters which reveals filtering operators reminiscent
of classic unsharp masking filters and anisotropic diffusion along boundaries.

To summarize our contribution: (1) we propose a novel, end-to-end
trainable, learning framework for solving various low-level image
reconstruction tasks;
(2) we show this framework is highly interpretable and controllable, enabling
direct post-hoc analysis of how the reconstructed image is generated from
the degraded input;
(3) we show experimentally that predictive filter flow outperforms the
state-of-the-art methods remarkably on the three different tasks, non-uniform
motion blur removal, compression artifact reduction and single image
super-resolution.


\section{Related Work}

Our work is inspired by filter flow~\cite{seitz2009filter}, which is an
optimization based method for finding a linear transformation relating nearby
pixel values in a pair of images.  By imposing additional constraints on
certain structural properties of these filters, it serves as a general
framework for understanding a wide variety of low-level vision problems.
However, filter flow as originally formulated has some obvious shortcomings.
First, it requires prior knowledge to specify a set of constraints needed to
produce good results.  It is not always straightforward to model or even come
up with such knowledge-based constraints.  Second, solving for an optimal
filter flow is compute intensive; it may take up to 20 hours to compute over a
pair of 500$\times$500 images~\cite{seitz2009filter}.  We address these by
directly predicting flows from image data.  We leverage predictive filter flow
for targeting three specific image reconstruction tasks which can be framed as
performing spatially variant filtering over local image patches.

%

\noindent {\bf Non-Uniform Blind Motion Blur Removal}
is an extremely challenging yet practically significant task of removing blur
caused by object motion or camera shake on a blurry photo. The blur kernel
is unknown and may vary over the image.  Recent methods estimate blur kernels
locally at patch level, and adopt an optimization method for deblurring the
patches~\cite{sun2015learning, bahat2017non}.
\cite{whyte2012non,hradivs2015convolutional,sun2015learning}
leverage prior information
about smooth motion by selecting from a predefine discretized set of linear
blur kernels.  These methods are computationally expensive as an iterative
solver is required for deconvolution after estimating the blur
kernel~\cite{cho2011handling};
and the deep learning approach cannot generalize well to novel motion
kernels~\cite{xu2014deep,sun2015learning,hradivs2015convolutional,schuler2016learning}.

\noindent{\bf Compression Artifact Reduction} is of significance as lossy image
compression is ubiquitous for reducing the size of images transmitted over the
web and recorded on data storage media. However, high compression rates come
with visual artifacts that degrade the image quality and thus user experience.
Among various compression algorithms, JPEG has become the most widely accepted
standard in lossy image compression with several (non-invertible)
transforms~\cite{wallace1992jpeg}, i.e., downsampling and DCT quantization.
Removing artifacts from jpeg compression can be viewed as a practical variant
of natural image denoising problems~\cite{buades2005non, jain2009natural}.
Recent methods based on deep convolutional neural networks trained to take as
input the compressed image and output the denoised image directly achieve good
performance~\cite{dong2015compression,svoboda2016compression,cavigelli2017cas}.

\noindent {\bf Single Image Super-Resolution} aims at recovering a
high-resolution image from a single low-resolution image. This problem is
inherently ill-posed as a multiplicity of solutions exists for any given
low-resolution input.  Many methods adopt an example-based
strategy~\cite{yang2014single} requiring an optimization solver, others are
based on deep convolutional neural nets~\cite{dong2016image,ledig2017photo}
which achieve the state-of-the-art and real-time performance.  The deep
learning methods take as input the low-resolution image (usually 4$\times$
upsampled one using bicubic interpolation), and output the high-resolution
image directly.

\section{Predictive Filter Flow}
Filter flow models image transformations $\I_1 \rightarrow \I_2$ as a linear
mapping where each output pixel only depends on a local neighborhood of the
input. Find such a flow can be framed as solving a constrained linear system
\begin{equation}
\I_2 = {\bf T} \I_1, \quad {\bf T} \in {\bf\Gamma}.
\label{eq:filterflow}
\end{equation}
where $\bf T$ is a matrix whose rows act separately on a vectorized
version of the source image $\I_1$.
For the model~\ref{eq:filterflow} to make sense,
${\bf T} \in {\bf\Gamma}$ must serve as a placeholder for the entire set of
additional constraints on the operator which enables a unique solution
that satisfies our expectations for particular problems of interest.
For example, standard convolution corresponds to $\bf T$ being a circulant
matrix whose rows are cyclic permutations of a single set of filter weights
which are typically constrained to have compact localized non-zero support.
For a theoretical perspective, Filter Flow model~\ref{eq:filterflow} is simple
and elegant, but directly solving Eq.~\ref{eq:filterflow} is intractable for
image sizes we typically encounter in practice, particularly when the filters
are allowed to vary spatially.

\subsection{Learning to predict flows}
Instead of optimizing over $\bf T$ directly,
we seek for a learnable function $f_{\w}(\cdot)$ parameterized by $\w$
that predicts the transformation $\bf \hat T$ specific to image $\I_1$
taken as input:
\begin{equation}
\I_2 \approx { \bf \hat T} \I_1, \quad {\bf {\hat T}}\equiv f_{\w}(\I_1),
\label{eq:predictivefilterflow1}
\end{equation}
We call this model Predictive Filter Flow.  Manually designing such a function
$f_{\w}(\cdot)$ isn't feasible in general, therefore we learn a specific $f_{\w}$
under the assumption that $\I_1,\I_2$ are drawn from some fixed joint
distribution.

Given sampled image pairs, $\{(\I_1^i, \I_2^i)\}$, where $i=1,\dots, N$, we seek
parameters $\w$ that minimize the difference between a recovered image
$\hat\I_2$ and the real one $\I_2$ measured by some loss $\ell$.
\begin{equation}
\begin{split}
& \min_{\w} \sum_{i=1}^{N} \ell( \I_2^i - f_{\w}(\I_1^i)\cdot\I_1^i ) + \mathcal{R}(f_{\w}(\I_1^i)), \\
& \text{s.t.  constraint on } \w
\end{split}
\label{eq:predictivefilterflow}
\end{equation}
Note that constraints on $\w$ are different from constraints ${\bf\Gamma}$ used in
Filter Flow.  In practice, we enforce hard constraints via our choice of the
architecture/functional form of $f$ along with soft-constraints via additional
regularization term $\mathcal{R}$. We also adopt commonly used $L_2$
regularization on $\w$ to reduce overfitting. There are a range of possible
choices for measuring the difference between two images. In our experiments, we
simply use the robust $L_1$ norm to measure the pixel-level difference.

\paragraph{Filter locality} In principle, each pixel output $\I_2$
in Eq.~\ref{eq:predictivefilterflow} can depend on all input pixels $\I_2$. We
introduce the structural constraint that each output pixel only depends on a
corresponding local neighborhood of the input.  The size of this neighborhood
is thus a hyper-parameter of the model. We note that while the predicted
filter flow ${\bf {\hat T}}$ acts locally, the estimation of the correct local
flow within a patch can depend on global context captured by large receptive
fields in the predictor $f_\w(\cdot)$.

In practice, this constraint is implemented by using the ``im2col'' operation
to vectorize the local neighborhood patch centered at each pixel and compute
the inner product of this vector with the corresponding predicted filter.  This
operation is highly optimized for available hardware architectures in most deep
learning libraries and has time and space cost similar to computing a single
convolution. For example, if the filter size is 20$\times$20, the last layer of
the CNN model $f_\w(\cdot)$ outputs a three-dimensional array with a channel
dimension of $400$, which is comparable to feature activations at a single
layer of typical CNN
architectures~\cite{krizhevsky2012imagenet,simonyan2014very,he2016deep}.


\paragraph{Other filter constraints}
Various priori constraints on the filter flow ${\bf \hat T}\equiv f_{\w}(\I_1)$
can be added easily to enable better model training.
For example, if smoothness is desired, an $L_2$ regularization on the (1st
order or 2nd order) derivative of the filter flow maps can be inserted during
training; if sparsity is desired, an $L_1$ regularization on the filter flows
can be added easily.  In our work, we add sum-to-one and non-negative
constraints on the filters for the task of non-uniform motion blur removal,
meaning that the values in each filter should be non-negative and sum-to-one by
assuming there is no lighting change.  This can be easily done by inserting a
softmax transform across channels of the predicted filter weights.
For other tasks, we simply let the model output free-form filters with
no further constraints on the weights.


\paragraph{Self-Supervision}
Though the proposed framework for training Predictive Filter Flow requires
paired inputs and target outputs, we note that generating training data for
many reconstruction tasks can be accomplished automatically without manual
labeling.  Given a pool of high quality images, we can automatically generate
low-resolution, blurred or JPEG degraded counterparts to use in training
(see Section~\ref{sec:exp}).  This can also be generalized to so-called
self-supervised training for predicting flows between video frames or
stereo pairs.


\subsection{Model Architecture and Training}

\begin{figure*}[ht]
    \centering
    \begin{minipage}{0.995\textwidth}
        \centering
        \includegraphics[width=1\linewidth]{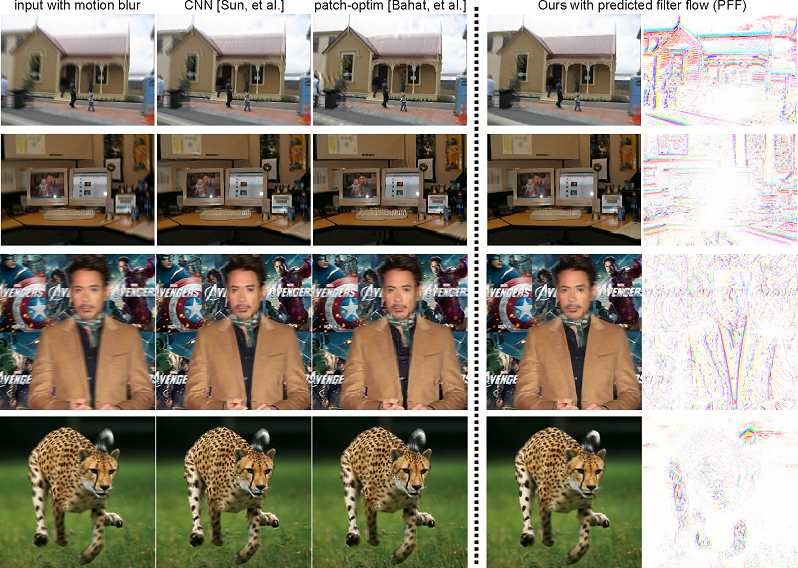}
        \captionsetup{width=0.999\textwidth}
    \end{minipage}
    \vspace{-3mm}
\caption{Visual comparison of our method (\emph{PFF}) to \emph{CNN [Sun, et al.]}~\cite{sun2015learning} and
    \emph{patch-optim [Bahat, et al.]}~\cite{bahat2017non}
    on testing images released by~\cite{bahat2017non}. Please be guided with the strong edges in the
    filter flow maps to compare visual details in the deblurred images by different methods.
Also note that the bottom two rows display images from the real-world,
meaning they are not synthesized and there is no blur ground-truth for them.
Best view in color and zoom-in.}
\label{fig:demo_motion_deblur}
\vspace{-3mm}
\end{figure*}

Our basic framework is largely agnostic to the choice of architectures,
learning method, and loss functions.  In our experiments, we utilize to a
two-stream architecture as shown in Fig.~\ref{fig:diagram}.  The first stream
is a simple 18-layer network with 3$\times$3 convolutional layers, skip
connections~\cite{he2016deep}, pooling layers and upsampling layers; the second
stream is a shallow but full-resolution network with no pooling.
The first stream has larger receptive fields for estimating per-pixel filters
by considering long-range contextual information, while the second stream keeps
original resolution as input image without inducing spatial information loss.
Batch normalization~\cite{ioffe2015batch} is also inserted between a convolution
layer and ReLU layer~\cite{nair2010rectified}.
The Predictive Filter Flow is self-supervised so we could generate an unlimited
amount of image pairs for training very large models. However, we find a light-weight
architecture trained over moderate-scale training set performs quite well.
Since our architecture is different from other feed-forward
image-to-image regression CNNs, we also report the baseline performance of the
two-stream architecture trained to directly predict the reconstructed image
rather than the filter coefficients.

For training, we crop 64$\times$64-resolution patches to form a batch of size
56. Since the model adapts to patch boundary effects seen during training, at
test time we apply it to non-overlapping tiles of the input image.  However, we
note that the model is fully convolutional so it could be trained over larger
patches to avoid boundary effects and applied to arbitrary size inputs.

We use ADAM optimization method during training~\cite{kingma2014adam},
with initial learning 0.0005 and coefficients 0.9 and 0.999
for computing running averages of gradient and its square. As for the training loss,
we simply use the $\ell_1$-norm loss measuring absolute difference over pixel intensities.
We train our model from scratch on a single NVIDIA TITAN X GPU,
and terminate after several hundred epochs\footnote{Models with early
termination ($\sim$2 hours for dozens of epochs) still achieve very good
performance, but top performance appears after 1--2 days training.
The code and models can be found in
\url{https://github.com/aimerykong/predictive-filter-flow}}.

\section{Experiments}
\label{sec:exp}

We evaluate the proposed Predictive Filter Flow framework (PFF) on three
low-level vision tasks: non-uniform motion blur removal, JPEG compression
artifact reduction and single image super-resolution.  We first describe the
datasets and evaluation metrics, and then compare with state-of-the-art methods
on the three tasks in separate subsections, respectively.

\subsection{Datasets and Metrics}
We use the high-resolution images in DIV2K dataset~\cite{Agustsson_2017_CVPR_Workshops}
and BSDS500 training set~\cite{martin2001database} for training all our models on the three tasks.
This results into a total of 1,200 training images.
We evaluate each model over different datasets specific to the task.
Concretely,
we test our model for non-uniform motion blur removal over the dataset introduced in
\cite{bahat2017non},
which contains large motion blur up to 38 pixels.
We evaluate over the classic LIVE1 dataset~\cite{wang2004image} for JPEG compression artifacts reduction,
and Set5~\cite{bevilacqua2012low} and Set14~\cite{zeyde2010single} for single image super-resolution.

To quantitatively measure performance, we use Peak-Signal-to-Noise-Ratio (PSNR)
and Structural Similarity Index
(SSIM)~\cite{wang2004image} over the Y channel in YCbCr color space between
the output quality image and the original image.
This is a standard practice in literature for quantitatively measuring the recovered image quality.

\subsection{Non-Uniform Motion Blur Removal}

{
\setlength{\tabcolsep}{0.9em} 
\begin{table}[t]
\centering
\caption{Comparison on motion blur removal over the non-uniform
motion blur dataset~\cite{bahat2017non}. For the two metrics,
the larger value means better performance of the model. }
\vspace{-1mm}
\small
\begin{tabular}{l c c c c c c c c c c c c c} 
\hline\hline
    & \multicolumn{5}{c} {\texttt{Moderate Blur}}  \\
    \cmidrule(r){2-6}
    metric & \cite{xu2013unnatural} &  \cite{sun2015learning} &  \cite{bahat2017non} & CNN & {\bf PFF} \\
    \hline
PSNR        & 22.88 & 24.14 & 24.87 & 24.51 & {\bf 25.39}  \\
SSIM        & 0.68  & 0.714 & 0.743 & 0.725 & {\bf 0.786} \\
\hline
&  \multicolumn{5}{c} {\texttt{Large Blur}}    \\
\cmidrule(r){2-6}
    metric & \cite{xu2013unnatural} &  \cite{sun2015learning} &  \cite{bahat2017non} & CNN & {\bf PFF} \\
    \hline
PSNR        & 20.47 & 20.84 & 22.01 & 21.06 & {\bf 22.30} \\
SSIM        & 0.54  & 0.56  & 0.624 & 0.560 & {\bf 0.638} \\
\hline\hline
\end{tabular}
\label{tab:compare_motion_deblur}
\end{table}
}

To train models for non-uniform motion blur removal, we generate the
64$\times$64-resolution blurry patches from clear ones using random linear
kernels~\cite{sun2015learning}, which are of size 30$\times$30 and have motion
vector with random orientation in $[0, 180^\circ]$ degrees and random length in
$[1,30]$ pixels.  We set the predicted filter size to be 17$\times$17 so the
model outputs 17$\times$17$=$289 filter weights at each image location.  Note
that we generate training pairs on the fly during training, so our model can
deal with a wide range of motion blurs.  This is advantageous over methods
in~\cite{sun2015learning,bahat2017non} which require a predefined set of blur
kernels used for deconvolution through some offline algorithm.

In Table~\ref{tab:compare_motion_deblur}, we list the comparison with the
state-of-the-art methods over the released test set by~\cite{bahat2017non}.
There are two subsets in the dataset, one with moderate motion blur and the
other with large blur.  We also report our CNN models based on the proposed
two-stream architecture that outputs the quality images directly by taking as
input the blurry ones.  Our CNN model outperforms the one in
\cite{sun2015learning} which trains a CNN for predicting the blur kernel over a
patch, but carries out non-blind deconvolution with the estimated kernel for
the final quality image.  We attribute our better performance to two reasons.
First, our CNN model learns a direct inverse mapping from blurry patch to its
clear counterpart based on the learned image distribution,
whereas \cite{sun2015learning} only estimates the blur kernel for the patch and uses
an offline optimization for non-blind deblurring, resulting in some artifacts
such as ringing.  Second, our CNN architecture is higher fidelity than the one
used in \cite{sun2015learning}, as ours outputs full-resolution result and
learns internally to minimize artifacts, e.g., aliasing and ringing effect.

\begin{figure*}[ht]
    \centering
    \begin{minipage}{0.995\textwidth}
        \centering
        \includegraphics[width=1\linewidth]{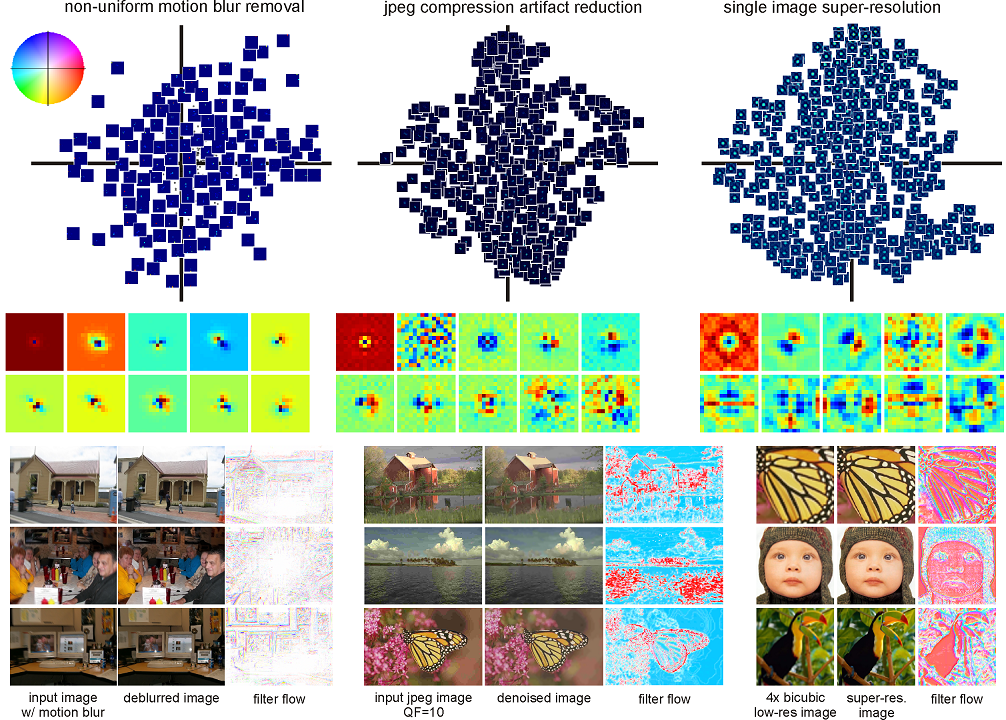}
        \captionsetup{width=0.999\textwidth}
    \end{minipage}
    \vspace{-3mm}
\caption{Three row-wise panels:
(1) We run K-means ($K$=400) on all filters synthesized by the model over the test set,
and visualize the 400 centroid kernels using t-SNE on a 2D plane;
(2) top ten principal components of the synthesized filters;
(3) visualizing the color coded filter flow along with input and quality
image. Each pixels filter is assigned to the nearest centroid and the color for the
centroid is based on the 2D t-SNE embedding using the color chart shown at top left.}
\label{fig:tSNE_filter_visualization}
\vspace{-1mm}
\end{figure*}

From the table, we can see our PFF model outperforms all the other methods by
a fair margin.  To understand where our model performs better,
we visualize the qualitative results in Fig.~\ref{fig:demo_motion_deblur},
along with the filter flow maps as output from PFF. We can't easily visualize
the 289 dimensional filters.  However, since the predicted weights $\hat T$ are
positive and $L_1$ normalized, we can treat them as a distribution which we
summarize by computing the expected flow vector
\[
\begin{bmatrix}
v_x(i,j)\\ v_y(i,j)
\end{bmatrix}
= \sum_{x,y} {\hat T}_{ij,xy}
\begin{bmatrix}
x - i\\ y - j
\end{bmatrix}
\]
where $ij$ is a particular output pixel and $xy$ indexes the input pixels.
This can be interpreted as the optical flow (delta filter) which most closely
approximates the predicted filter flow.  We use the the color legend shown in
top-left of Fig.~\ref{fig:tSNE_filter_visualization}.

The last two rows of Fig.~\ref{fig:demo_motion_deblur} show the results over
real-world blurry images for which there is no ``blur-free'' ground-truth.  We
can clearly see that images produced by PFF have less artifacts such as ringing
artifacts around sharp edges~\cite{sun2015learning,bahat2017non}.
Interestingly, from the filter flow maps, we can see that the expected flow
vectors are large near high contrast boundaries and smaller in regions that are
already in sharp focus or which are uniform in color.

Although we define the filter size as 17$\times$17, which is much smaller than
the maximum shift in the largest blur (up to 30 pixels), our model still
handles large motion blur and performs better than~\cite{bahat2017non}.  We
assume it should be possible to utilize larger filter sizes but we did not
observe further improvements when training models to synthesize larger
per-pixel kernels. This suggests that a larger blurry dataset is needed to
validate this point in future work.

We also considered an iterative variant of our model in which we feed the
resulting deblurred image back as input to the model.  However, we found
relatively little improvement with additional iterations (results shown in the
appendix). We conjecture that, although the model was trained
with a wide range of blurred examples, the statistics of the transformed image
from the first iteration are sufficiently different than the blurred training
inputs.  One solution could be inserting adversarial loss to push the model to
generate more fine-grained textures (as done in~\cite{ledig2017photo} for image
super-resolution).

\begin{figure}[t]
    \centering
    \begin{minipage}{0.497\textwidth}
        \centering
        \includegraphics[width=1\linewidth]{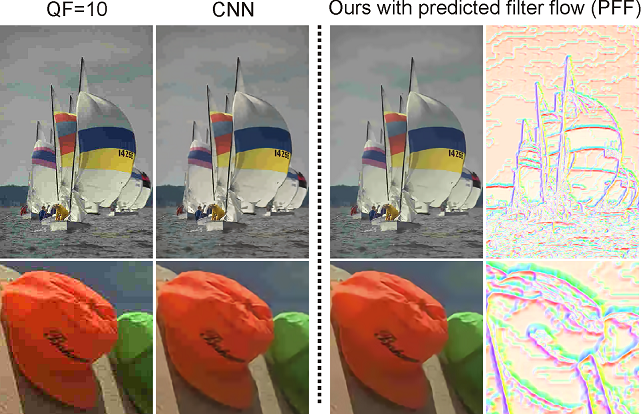}
    \end{minipage}
    \vspace{-2mm}
    \caption{Visual comparison of our methods (PFF and CNN). Strong edges
    in the expected flow map (right) highlight areas where most apparent
    artifacts are removed.  More results can be found in the appendix.  
    Best viewed in color and zoomed-in.}
    \label{fig:demo_jpeg}
    \vspace{-1mm}
\end{figure}

\subsection{JPEG Compression Artifact Reduction}

{
\setlength{\tabcolsep}{0.26em} 
\begin{table}[t]
\centering
\caption{Comparison on JPEG compression artifact reduction over LIVE1 dataset~\cite{wang2004image}.
PSNR and SSIM are used as metrics listed on two rows respectively in each macro row grid
(the larger the better). }
\vspace{-1mm}
\footnotesize
\begin{tabular}{l | c c c c c c c c c c c c c} 
\hline
QF      & JPEG &  SA-DCT &  AR-CNN & L4 & CAS-CNN & MWCNN  & {\bf PFF} \\
& & \cite{foi2007pointwise} &  \cite{dong2015compression} & \cite{svoboda2016compression}
 & \cite{cavigelli2017cas}  &  \cite{liu2018multi} & \\
\hline
\multirow{2}{*}{10}
        & 27.77 & 28.65 & 29.13     & 29.08 & 29.44  & 29.69  & {\bf 29.82}  \\
        & 0.791 & 0.809 & 0.823     & 0.824 & 0.833  & 0.825  & {\bf 0.836}  \\
\hline
\multirow{2}{*}{20}
        & 30.07 & 30.81 & 31.40     & 31.42 & 31.70  & 32.04  & {\bf 32.14}  \\
        & 0.868 & 0.878 & 0.890     & 0.890 & 0.895  & 0.889  & {\bf 0.905}  \\
\hline
\multirow{2}{*}{40}
        & 32.35 & 32.99 & 33.63     & 33.77 & 34.10  & 34.45  & {\bf 34.67}  \\
        & 0.917 & 0.940 & 0.931     & ---   & 0.937  & 0.930  & {\bf 0.949}  \\
\hline
\end{tabular}
\label{tab:compare_motion_deblur2}
\end{table}
}

Similar to training for image deblurring,
we generate JPEG compressed image patches from original non-compressed ones
on the fly during training.
This can be easily done using JPEG compression function by varying the
quality factor (QF) of interest.

\begin{figure}[t]
    \centering
    \begin{minipage}{0.5\textwidth}
        \centering
        \includegraphics[width=1\linewidth]{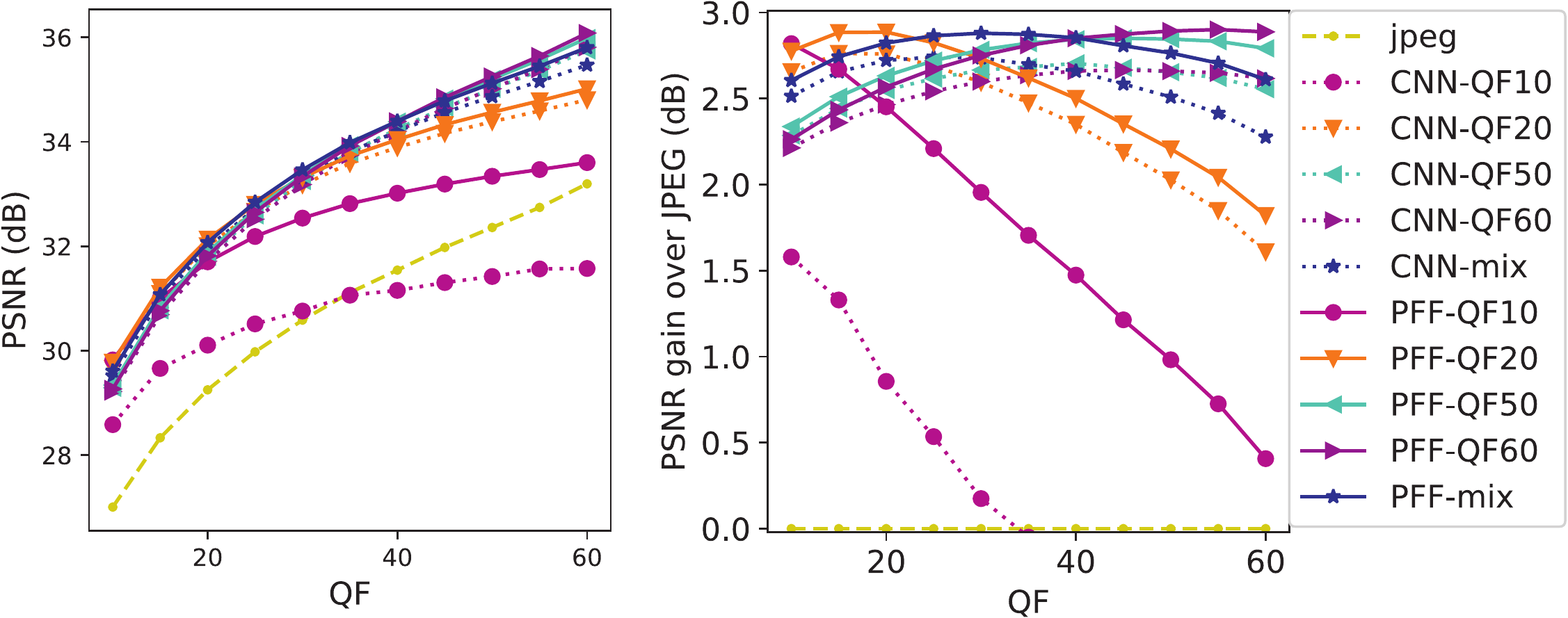}
        \vspace{-8mm}
        \caption*{PSNR improvements.}
    \end{minipage}%
    \vspace{3mm}
    \begin{minipage}{.5\textwidth}
        \centering
        \includegraphics[width=1\linewidth]{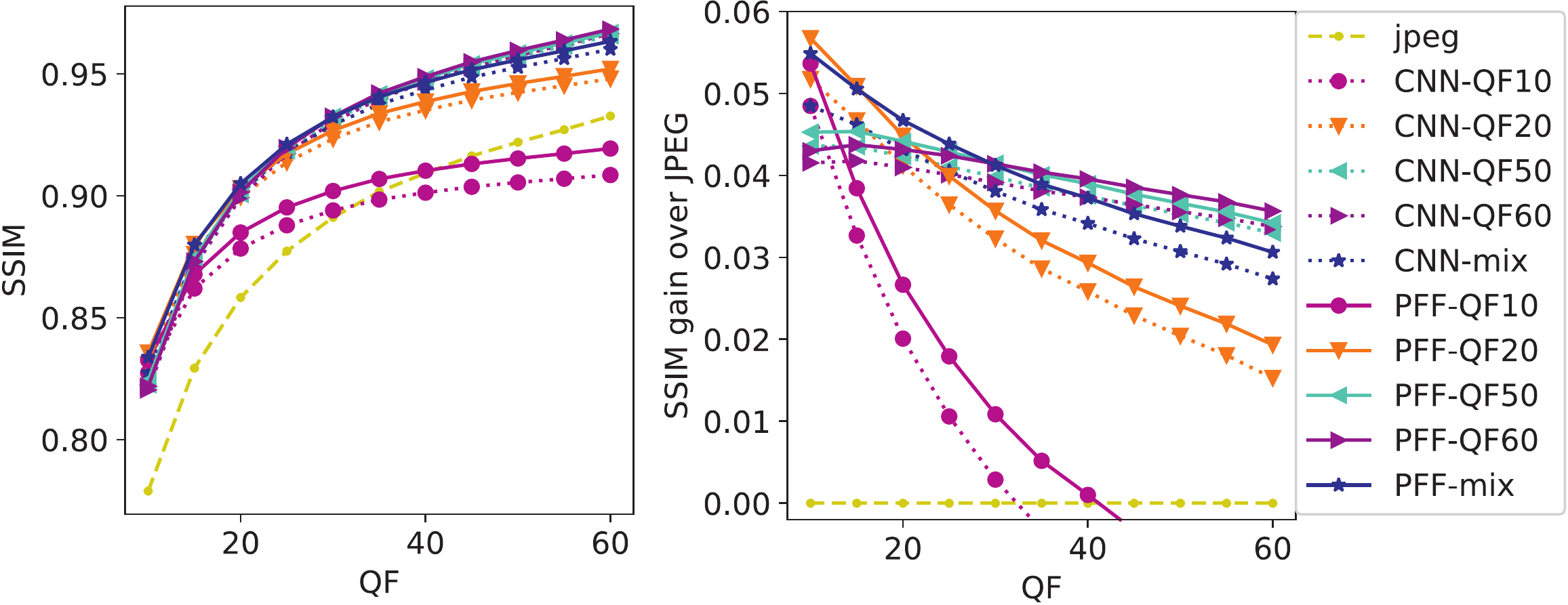}
        \vspace{-8mm}
        \caption*{SSIM improvements.}
    \end{minipage}
    \vspace{-4mm}
    \caption{Performance vs. training data with different compression quality
    factors measured by PSNR and SSIM and their performance gains, over the
    LIVE1 dataset.  The original JPEG compression is plotted for baseline.}
    \label{fig:curves_jpeg}
\end{figure}

In Table~\ref{tab:compare_motion_deblur2},
we list the performance of our model and compare to the state-of-the-art methods.
We note that our final PFF achieves the best among all the methods.
Our CNN baseline model also achieves on-par performance with state-of-the-art,
though we do not show in the table, we draw the performance under the ablation
study in Fig.~\ref{fig:curves_jpeg}.
Specifically, we study how our model trained with single or a mixed QFs affect
the performance when tested on image compressed with a range of different QFs.
We plot the detailed performances of our CNN and PFF in terms of
absolute measurements by PSNR and SSIM, and the increase in PSNR between
the reconstructed and JPEG compressed image.

We can see that, though a model trained with QF=10 overfits the dataset, all
the other models achieve generalizable and stable performance.  Basically, a
model trained on a single QF brings the largest performance gain over images
compressed with the same QF.  Moreover, when our model is trained with mixed
quality factors, its performance is quite stable and competitive with
quality-specific models across different compression quality factors.  This
indicates that our model is of practical value in real-world applications.

In Fig.~\ref{fig:demo_jpeg}, we  demonstrate qualitative comparison between
CNN and PFF.  The output filter flow maps indicate from the colorful edges how
the pixels are warped from the neighborhood in the input image.  This also
clearly shows where the JPEG image degrades most, e.g., the large sky region is
quantized by JPEG compression.  Though CNN makes the block effect smooth to
some extent, our PFF produces the best visual quality, smoothing the block
artifact while maintaining both high- and low-frequency details.

\begin{figure}[t]
    \centering
    \begin{minipage}{0.497\textwidth}
        \centering
        \includegraphics[width=1\linewidth]{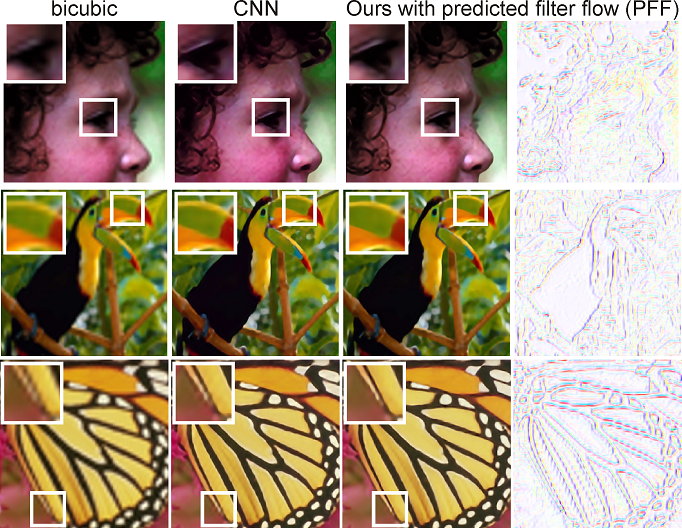}
    \end{minipage}
    \vspace{-1mm}
    \caption{Visual comparison of our method (PFF) to CNN, each image is super-resolved
    ($\times$4).  More results can be found in the appendix.
        Best view in color and zoom-in.}
    \label{fig:demo_SISR}
    \vspace{-3mm}
\end{figure}

\subsection{Single Image Super-Resolution}
In this work,
we only generate pairs to super-resolve images 4$\times$ larger.
To generate training pairs,
for each original image,
we downsample $\frac{1}{4}$$\times$ and upsample 4$\times$
again using bicubic interpolation (with anti-aliasing).
The 4$\times$ upsampled image from the low-resolution is the input to our model.
Therefore,
a super-resolution model is expected to be learned for sharpening the input image.

In Table~\ref{tab:eval_SISR},
we compare our PFF model quantitatively with other methods.
We can see that our model outperforms the others on both test sets.
In Fig.~\ref{fig:demo_SISR},
we compare visually over bicubic interpolation,
CNN and PFF. We can see from the zoom-in regions that our PFF generates sharper
boundaries and delivers an anti-aliasing functionality.
The filter flow maps once again act as a guide,
illustrating where the smoothing happens and where sharpening happens.
Especially, the filter maps demonstrate from the strong colorful
edges where the pixels undergo larger transforms.
In next section,
we visualize the per-pixel kernels to have an in-depth understanding.

{
\setlength{\tabcolsep}{0.4em} 
\begin{table}
\caption{Comparison on single image super-resolution ($\times$4) over the classic
Set5~\cite{bevilacqua2012low} and Set14~\cite{zeyde2010single} datasets.
The metrics used here are PSNR (dB) and SSIM listed as two rows, respectively.
}
\vspace{-3mm}
\footnotesize
\begin{center}
\begin{tabular}{ c| c c c c c c c c c c}
\hline
     & Bicubic & NE+LLE & KK  & A+ & SRCNN & RDN+ & {\bf PFF} \\
     &   & \cite{chang2004super} & \cite{kim2010single}
    & \cite{timofte2013anchored} & \cite{dong2016image}  & \cite{zhang2018residual} &   \\
\hline
\multirow{2}{*}{\rotatebox[origin=c]{90}{{\centering  {\scriptsize Set5}}}}
&  28.42  &  29.61  & 29.69  & 30.28  & 30.49  & 32.61 & \textbf{32.74} \\
&  0.8104 &  0.8402 & 0.8419 & 0.8603 & 0.8628 & 0.9003 & \textbf{0.9021} \\
\hline
\multirow{2}{*}{\rotatebox[origin=c]{90}{{\centering {\scriptsize Set14}}}}
&  26.00  &  26.81  & 26.85  & 27.32  & 27.50  & 28.92 & \textbf{28.98} \\
&  0.7019 &  0.7331 & 0.7352 & 0.7491 & 0.7513 & 0.7893 & \textbf{0.7904} \\
\hline
\end{tabular}
\end{center}
\label{tab:eval_SISR}
\vspace{-0.5mm}
\end{table}
}

\section{Visualization and Analysis}
We explored a number of techniques to visualize the predicted filter flows for
different tasks.  First, we ran k-means on predicted filters from the set of
test images for each the three tasks, respectively, to cluster the kernels into
$K$=400 groups.  Then we run t-SNE~\cite{maaten2008visualizing} over the 400
mean filters to display them in the image plane, shown by the scatter plots in
top row of Fig.~\ref{fig:tSNE_filter_visualization}.  Qualitative inspection
shows filters that can be interpreted as performing translation or integration
along lines of different orientation (non-uniform blur),  filling in
high-frequency detail (jpeg artifact reduction) and deformed Laplacian-like
filters (super-resolution).

We also examined the top 10 principal components of the predicted filters
(shown in the second row grid in Fig.~\ref{fig:tSNE_filter_visualization}).
The 10D principal subspace capture 99.65\%, 99.99\% and 99.99\% of the filter energy
for non-uniform blur, artifact removal and super resolution respectively.  PCA
reveals smooth, symmetric harmonic structure for super-resolution with some
intriguing vertical and horizontal features.

Finally, in order to summarize the spatially varying structure of the filters,
we use the 2D t-SNE embedding to assign a color to each centroid (as given by
the reference color chart shown top-left), and visualize the nearest centroid
for the filter at each filter location in the third row grid in
Fig.~\ref{fig:tSNE_filter_visualization}.  This visualization demonstrates
the filters as output by our model generally vary smoothly over the image
with discontinuities along salient edges and textured regions reminiscent
of anisotropic diffusion or bilateral filtering.

In summary, these visualizations provide a transparent view of how each
reconstructed pixel is assembled from the degraded input image.  We view this
as a notable advantage over other CNN-based models which simply perform
image-to-image regression. Unlike activations of intermediate layers of a CNN,
linear filter weights have a well defined semantics that can be visualized and
analyzed using well developed tools of linear signal processing.

\section{Conclusion and Future Work}
We propose a general, elegant and simple framework called Predictive Filter
Flow, which has direct applications to a broad range of image reconstruction
tasks.  Our framework generates space-variant per-pixel filters which are easy
to interpret and fast to compute at test time.  Through extensive experiments
over three different low-level vision tasks, we demonstrate this approach
outperforms the state-of-the-art methods.

In our experiments here, we only train light-weight models over patches,
However, we believe global image context is also important for these tasks
and is an obvious direction for future work.  For example, the global blur
structure conveys information about camera shake; super-resolution and
compression reduction can benefit from long-range interactions to reconstruct
high-frequency detail (as in non-local means). Moreover, we expect that the
interpretability of the output will be particularly appealing for interactive
and scientific applications such as medical imaging and biological microscopy
where predicted filters could be directly compared to physical models of the
imaging process.

\section*{Acknowledgement}
This project is supported by NSF grants IIS-1618806, IIS-1253538, DBI-1262547 and a hardware donation from NVIDIA.

{\small
\bibliographystyle{ieee}
\bibliography{egbib}
}

\newpage\newpage
\setcounter{section}{0}

\begin{center}
 {\large \textbf{Appendix}}
\end{center}

\emph{
In the supplementary material,
we first show more visualizations to understand the predicted filter flows,
then show if it is possible to refine the results
by iteratively feeding deblurred image to the same model
for the task of non-uniform motion blur removal.
We finally present more qualitative results for all the three tasks studied in this paper.
}

\section{Visualization of Per-Pixel Loading Factors}

\begin{figure}[t]
    \centering
    \begin{minipage}{0.49\textwidth}
        \centering
        \includegraphics[width=1\linewidth]{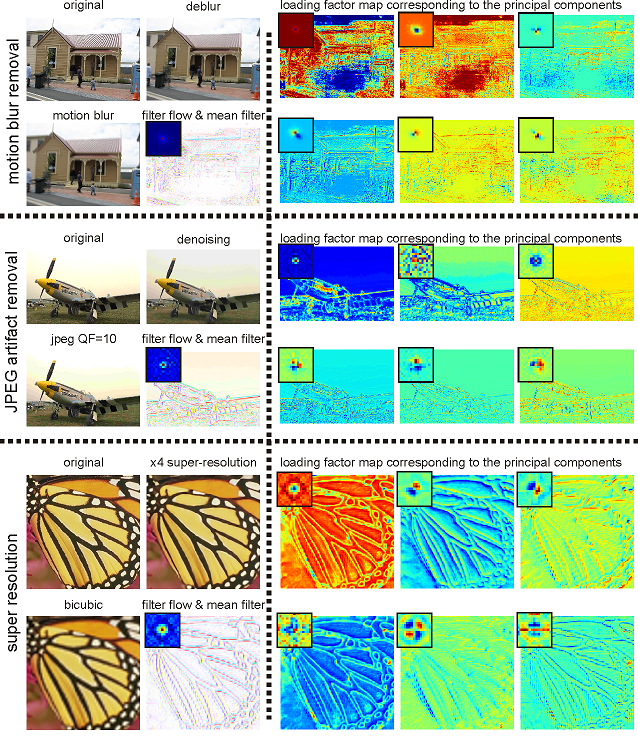}
    \end{minipage}
    \caption{
    We show the original image,
    low-quality input and the high-quality output by our model as well as the mean kernel and filter flow maps on the left panel,
    and the first six principal components and the corresponding loading factors as heatmap
    on the right panel.
    Best seen in color and zoom-in.}
    \label{fig:demo_loadingFactorPCA}
\end{figure}

As a supplementary visualization to the principal components by PCA shown in the main paper,
we can also visualize the per-pixel loading factors corresponding to each principal component.
We run PCA over testing set and show the first six principal components and the corresponding
per-pixel loading factors as a heatmap in Figure~\ref{fig:demo_loadingFactorPCA}.
With this visualization technique,
we can know what region has higher response to which component kernels.
Moreover,
given that the first ten principal components capture $\ge 99\%$ filter
energy (stated in the main paper),
we expect future work to predict compact per-pixel filters using low-rank technique,
which allows for incorporating long-range pixels through large predictive filters while
with compact features (thus memory consumption is reduced largely).

\section{Iteratively Removing Motion Blur }
\begin{figure*}[t]
    \centering
    \begin{minipage}{0.95\textwidth}
        \centering
        \includegraphics[width=1\linewidth]{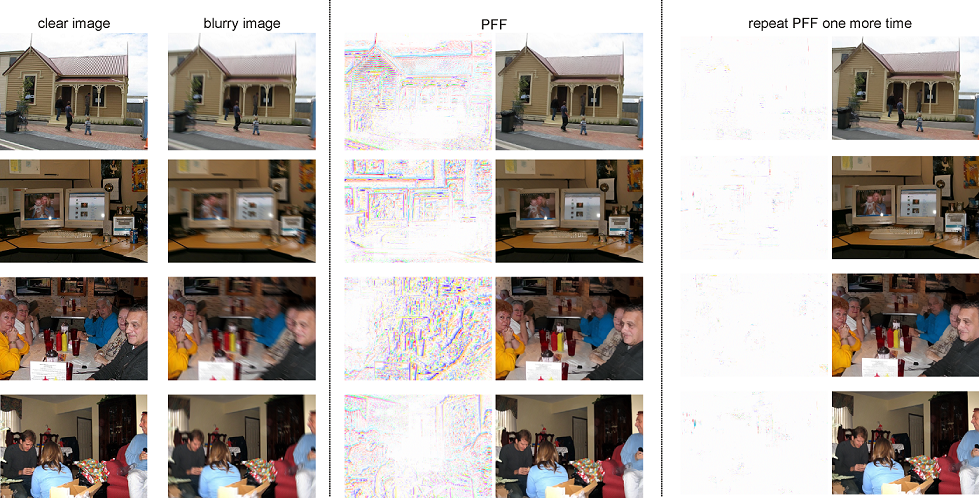}
    \end{minipage}
    \caption{
    We show deblurring results over some random  testing images from the dataset released by~\cite{bahat2017non}.
    We first feed the blurry images to PFF model, and obtain deblurred images;
    then we feed such deblurred images into the same PFF model again to see if this iterative practice refines
    the output.
    However,
    through the visualization that iteratively running the model changes very little as seen
    from the second filter flow maps.
    This helps qualitatively explain why iteratively running the model does not improve deblurring performance
    further.
    }
    \label{fig:iterativeRefine}
\end{figure*}

As the deblurred images are still not perfect,
we are interested in studying if we can improve performance by iteratively running the model, i.e.,
feeding the deblurred image as input to the same model one more time to get the result.
We denote this method as PFF+1.
Not much surprisingly,
we do not observe further improvement as listed in
Figure~\ref{tab:compare_motion_deblur_more}, instead,
such a practice even hurts performance slightly.
The qualitative results are shown in Figure~\ref{fig:iterativeRefine},
from which we can see the second run does not generate much change through the filter flow maps.
We believe the reason is that,
the deblurred images have different statistics from the original blurry input,
and the model is not trained with such deblurred images.
Therefore, it suggests two natural directions as future work for improving the results,
1) training explicitly with recurrent loops with multiple losses to improve the performance,
similar to~\cite{belagiannis2017recurrent,li2016iterative,romera2016recurrent,
kong2017recurrentSceneParsing,kong2018pixel},
or 2) simultaneously inserting an adversarial loss to force the model to hallucinate
details for realistic output, which can be useful in practice as done in~\cite{ledig2017photo}.

{
\setlength{\tabcolsep}{0.9em} 
\begin{table*}[t]
\centering
\caption{Comparison on motion blur removal over the non-uniform
motion blur dataset~\cite{bahat2017non}. PFF+1 means we perform PFF one more time by taking
as input the deblurred image by the same model. }
\vspace{-1mm}
\begin{tabular}{l c c c c c c c c c c c c c} 
\hline\hline
    & \multicolumn{6}{c} {\texttt{Moderate Blur}}  \\
    \cmidrule(r){2-7}
    metric & \cite{xu2013unnatural} &  \cite{sun2015learning} &  \cite{bahat2017non} & CNN
    & {\bf PFF}
    & PFF+1 \\
    \hline
PSNR        & 22.88 & 24.14 & 24.87 & 24.51 & {\bf 25.39}  & 25.28  \\
SSIM        & 0.68  & 0.714 & 0.743 & 0.725 & {\bf 0.786}  & 0.783  \\
\hline
&  \multicolumn{6}{c} {\texttt{Large Blur}}    \\
\cmidrule(r){2-7}
    metric & \cite{xu2013unnatural} &  \cite{sun2015learning} &  \cite{bahat2017non} & CNN
    & {\bf PFF}
    & PFF+1 \\
    \hline
PSNR        & 20.47 & 20.84 & 22.01 & 21.06 & {\bf 22.30}  & 22.21  \\
SSIM        & 0.54  & 0.56  & 0.624 & 0.560 & {\bf 0.638}  & 0.633  \\
\hline\hline
\end{tabular}
\label{tab:compare_motion_deblur_more}
\end{table*}
}

\section{More Qualitative Results}

\begin{figure*}[t]
    \centering
    \begin{minipage}{0.97\textwidth}
        \centering
        \includegraphics[width=1\linewidth]{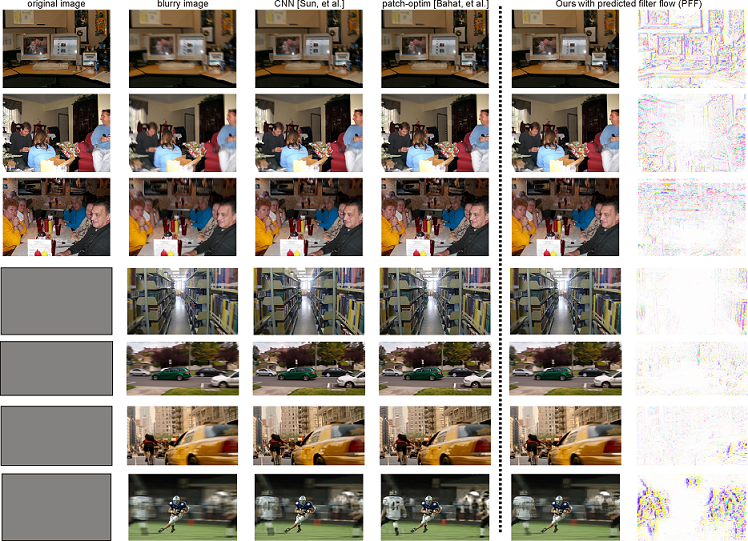}
    \end{minipage}
    \caption{Visual comparison of our method (\emph{PFF}) to \emph{CNN [Sun, et al.]}~\cite{sun2015learning} and
    \emph{patch-optim [Bahat, et al.]}~\cite{bahat2017non}
    on more testing images released by~\cite{bahat2017non}.
    Please be guided with the strong edges in the filter flow maps to compare visual details in the deblurred images by different methods.
    The last four rows show real-world blurry images without ``ground-truth'' blur.
    Note that for the last image,
    there is very large blur caused by the motion of football players.
    As our model is not trained on larger kernels which should be able to cover the size of blur,
    it does not perform as well as \emph{patch-optim [Bahat, et al.]}~\cite{bahat2017non}.
    But it is clear that our model generates sharp edges in this task.
    Best view in color and zoom-in.}
    \label{fig:more_motionblur}
\end{figure*}

\begin{figure*}[t]
    \centering
    \begin{minipage}{0.97\textwidth}
        \centering
        \includegraphics[width=1\linewidth]{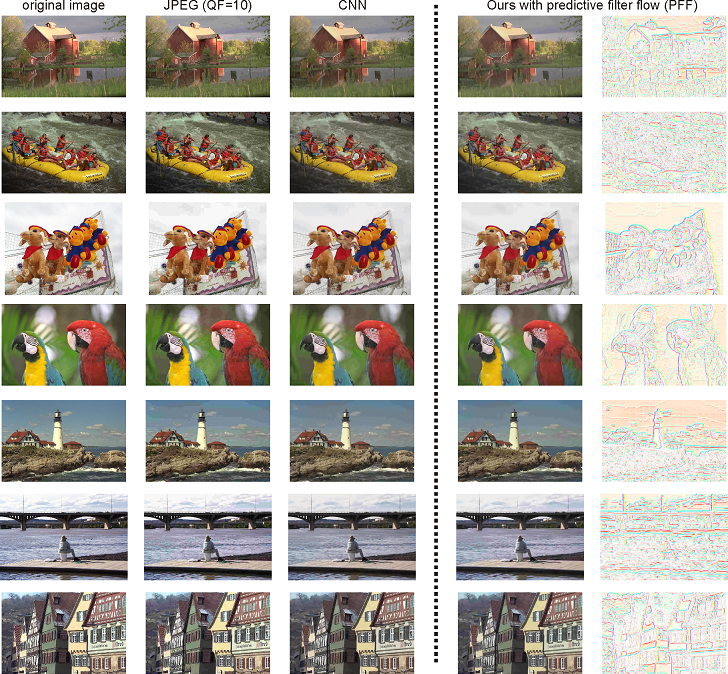}
    \end{minipage}
    \caption{
    Visual comparison between CNN and our method (\emph{PFF}) for JPEG compression artifact reduction.
    Here we compress the original images using JPEG method with quality factor (QF) as 10.
    Best view in color and zoom-in.}
    \label{fig:more_JPEG_reduction}
\end{figure*}

\begin{figure*}[t]
    \centering
    \begin{minipage}{0.77\textwidth}
        \centering
        \includegraphics[width=1\linewidth]{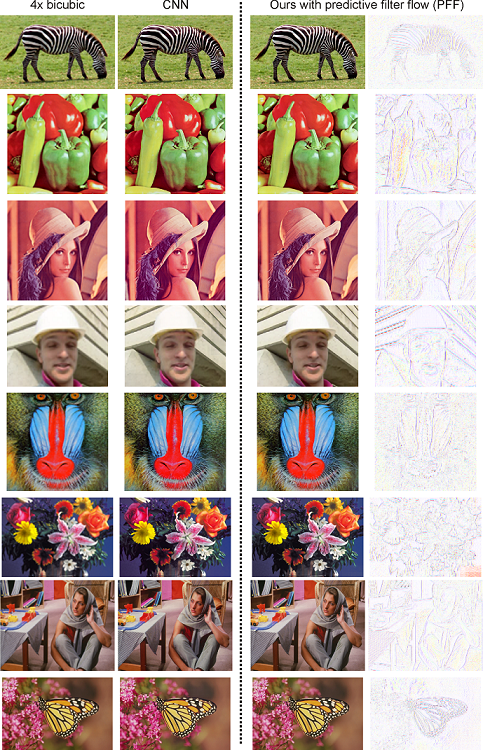}
    \end{minipage}
    \caption{
    Visual comparison between CNN and our method (\emph{PFF}) for single image super-resolution.
    Here all images are super-resolved by 4$\times$ larger.
    We show in the first column the results by bicubic interpolation.
    Best view in color and zoom-in.}
    \label{fig:more_SISR}
\end{figure*}

In Figure~\ref{fig:more_motionblur}, \ref{fig:more_JPEG_reduction} and \ref{fig:more_SISR},
we show more qualitative results for
non-uniform motion blur removal,
JPEG compression artifact reduction
and single image super-resolution, respectively.
From these comparisons and with the guide of filter flow maps,
we can see at what regions our PFF pays attention to and how it outperforms
the other methods.

\end{document}